\documentclass[pre,aps]{revtex4}
\usepackage{graphicx}

\begin{document}

\title{Magnetic solitons in frustrated ferromagnetic spin chain}
\author{D.~V.~Dmitriev}
\email{dmitriev@deom.chph.ras.ru}
\author{V.~Ya.~Krivnov}
\affiliation{Joint Institute of Chemical Physics of RAS, Kosygin
str. 4, 119334, Moscow, Russia.}
\date{}

\begin{abstract}
We study the classical anisotropic ferromagnetic spin chain with
frustration. The behavior of soliton and kink solutions in the
vicinity of the ground state phase transition from the ferromagnetic
to the spiral phase is studied. The dependence of the soliton energy
on small anisotropy parameter is established using scaling estimates
and numerical minimization of the energy functional. Conditions of
the existence of the solitons are determined. It is shown that
solitons survive in the spiral phase though with some restrictions
on their size. A comparison of the energies of the classical
solitons and the bound magnon complexes in the quantum model shows
the functional similarity between them. The influence of the
finite-size effects on the soliton states is studied and it is shown
that the localized solitons originate from the uniform state when
the system size exceeds some critical value depending on the
anisotropy.
\end{abstract}

\maketitle

\section{Introduction}

Lately, there has been considerable interest in low-dimensional spin
systems that exhibit frustration \cite{review}. A very interesting
class of such systems is chain compounds consisting of edge-sharing
$CuO_4$ units. Recently, a variety of these copper oxides were
synthesized and found to show unique physical properties
\cite{Mizuno,Masuda,Hase,Capogna,stefan05,stefan08}. The frustration
in these compounds arises from the competition of exchange
interactions between magnetic $Cu^{2+}$ ions carrying spins $1/2$.
Due to a specific geometry of these systems ($Cu-O-Cu$ angle is
close to $90^0$) the nearest-neighbor (NN) interaction is
ferromagnetic, while the next-nearest-neighbor (NNN) one is
antiferromagnetic and absolute values of these interactions are
comparable \cite{Mizuno,stefan08}. An appropriate model describing
the magnetic properties of such copper oxides is so-called F-AF spin
chain model the Hamiltonian of which has a form
\begin{equation}
H =
J_1\sum_{n=1}^N(S_n^xS_{n+1}^x+S_n^yS_{n+1}^y+\Delta_1S_n^zS_{n+1}^z)
+J_2\sum_{n=1}^N(S_n^xS_{n+2}^x+S_n^yS_{n+2}^y+\Delta_2S_n^zS_{n+1}^z)
\label{H}
\end{equation}
where $J_1<0$ and $J_2>0$.

This model is characterized by a frustration parameter
$\lambda=J_2/|J_1|$. The quantum F-AF $s=1/2$ model has been
intensively studied last years
\cite{Chubukov,KO,DK06,Vekua,Lu,DKR,Itoi,Kecke,Laeuchli}. Most of
studies of this model are related to the isotropic case
($\Delta_1=\Delta_2=1$). It is known that the ground state of the
isotropic model is ferromagnetic for $\lambda<1/4$. At $\lambda=1/4$
the phase transition to the incommensurate singlet phase with spiral
spin correlations takes place. Remarkably, this transition point
does not depend on spin value $s$ at $N\to\infty$. It was shown also
that the F-AF $s=1/2$ model with anisotropic interactions has a rich
phase diagram \cite{Somma}. In our papers \cite{DK08,DK09} we
investigated weakly anisotropic ($\Delta_1>1$, $\Delta_2=1$) quantum
spin-1/2 model (\ref{H}). It was shown that even small anisotropy
essentially effects on the properties of the model. In particular,
the transition point from the ferromagnetic to the spiral-like
ground state shifts from $\lambda=1/4$.

An interesting feature of the quantum anisotropic model (\ref{H}) is
the existence of the multimagnon bound complexes in the
ferromagnetic phase. These complexes govern the low-temperature
thermodynamics \cite{DK09,Johnson}. It was noted also that the NNN
interaction strongly affects the excitation spectrum especially in
the vicinity of the transition point $\lambda=1/4$.

It is known that there is a close relation between the multimagnon
bound complexes in the quantum spin models and soliton and kink
excitations in the classical counterparts. In particular, these
soliton states have been studied extensively for the classical
easy-axis ferromagnetic chain and a connection between them and the
bound magnon complexes in the quantum $s=1/2$ model (\ref{H}) at
$J_2=0$ was discussed \cite{Schneider}. On the other hand, the
influence of the frustration on the solitons has not been considered
before. The F-AF model represents a suitable model to study this
problem and to investigate the connection between the quantum
excitation spectrum and the soliton solutions of the classical
frustrated spin model.

In \cite{DK08} we showed that the ground state phase diagram of the
quantum model (\ref{H}) with both $\Delta_1\neq 1$ and $\Delta_2\neq
1$ are qualitatively similar to that for the model with the
anisotropy of the NN interaction only. Therefore, for simplicity we
consider model (\ref{H}) with $\Delta_1=\Delta>1$ and $\Delta_2=1$.
In this case model (\ref{H}) takes the form
\begin{equation}
H = -\sum\left(S_n^xS_{n+1}^x+S_n^yS_{n+1}^y +\Delta
S_n^zS_{n+1}^z-\Delta s^2\right)
+\lambda\sum(\mathbf{S}_{n}\cdot\mathbf{S}_{n+2}-s^2)  \label{HH}
\end{equation}
where we put $|J_1|$ as an energy unit and added constant shifts to
secure the energy of the ferromagnetic state to be zero.

The phase diagram of the classical F-AF model with $\Delta>1$
consists of the ferromagnetic and the spiral phases. The
ferromagnetic ground state is simple, i.e., it has all spins
parallel to the Z axis. However, the soliton excitations in this
phase are not trivial especially near the transition point between
the phases. Our main goal is to study the behavior of the solitons
in the vicinity of the isotropic transition (IT) point ($\Delta=1$,
$\lambda=1/4$) and to compare it with the properties of the
excitations of the quantum model. As it will be shown the
frustration effects strongly modify the soliton states especially
for small anisotropy. In particular, the exponent characterizing the
power dependence of the gap (the soliton energy with respect to the
ground state) on the anisotropy is different from that for
$\lambda=0$.

The paper is organized as follows. In Sec.II we represent the known
results for soliton solutions of the anisotropic ferromagnetic chain
(model (\ref{HH}) at $\lambda=0$). In Sec.III we consider the
classical continuum F-AF model in the vicinity of the ground state
phase transition from the ferromagnetic to the spiral phase. We
deduce the corresponding energy functional and establish the scaling
form of the soliton energy as a function of the anisotropy and the
frustration parameter. In this section we also obtain asymptotes of
the soliton solutions at large distances and determine the necessary
conditions of the soliton stability. In Sec.IV we present results of
the numerical minimization of the energy functional which confirm
the scaling estimations. In Sec.V we study the finite-size effects
on the soliton solution. In Sec.VI we show that the behavior of both
classic solitons and $m$-magnon quantum spin excitations are
functionally similar to $m$-boson bound complexes of the Bose model
with the attractive interaction if $m$ is not large. In Sec.VII we
give a summary of results.

\section{Classical model for $\lambda=0$ case}

In the classical approximation the spin operators $\mathbf{S}_n$ are
replaced by the classical vectors $\overrightarrow{S}_n$ of the
fixed length $s$ which are parameterized by spherical coordinates
\begin{equation}
\overrightarrow{S}_n = s(\cos\varphi_n\sin\theta_n,
\sin\varphi_n\sin\theta_n,\cos\theta_n)   \label{n}
\end{equation}

In terms of the angles $\theta_n$ and $\varphi_n$ the discrete
classical F-AF model (\ref{HH}) takes the form
\begin{eqnarray}
E &=& s^2\sum\left\{[1-\cos(\theta_{n+1}-\theta_n)]
-\lambda[1-\cos(\theta_{n+2}-\theta_n)]
+\alpha[1-\cos(\theta_{n+1})\cos(\theta_n)]\right\}  \nonumber \\
&& +s^2\sum\left\{
[1-\cos(\varphi_{n+1}-\varphi_n)]\sin\theta_n\sin\theta_{n+1}
-\lambda[1-\cos(\varphi_{n+2}-\varphi_n)]\sin\theta_n\sin\theta_{n+2}
\right\} \label{Edis}
\end{eqnarray}
where $\alpha=\Delta-1>0$.

In this Section we briefly review the known results for the
classical Heisenberg ferromagnetic chain with an easy-axis
anisotropy, i.e., model (\ref{Edis}) with $\lambda=0$. In this
case the model is exactly solved in the continuum limit
\cite{Steiner,KIK}.

The continuum limit of the classical model assumes that the vectors
$\overrightarrow{S}_n$ can be replaced by the classical vector field
$\overrightarrow{S}(x,t)$ with slowly varying orientations, so that
\begin{equation}
\overrightarrow{S}_{n+1}-\overrightarrow{S}_{n}\approx
\frac{\partial\overrightarrow{S}(x_n)}{\partial x}  \label{dSdx}
\end{equation}
where the lattice constant is chosen as unit length. The direction
of the vector field $\overrightarrow{S}(x,t)$ is determined by two
angular variables $\varphi(x,t)$ and $\theta(x,t)$ according to
Eq.(\ref{n}).

The dynamics of the vector field $\overrightarrow{S}(x,t)$ is
governed by the Landau-Lifshitz equation:
\begin{equation}
-\frac{\partial \overrightarrow{S}}{\partial t}
=\overrightarrow{S}\mathbf{\times }\frac{\delta E}{\delta
\overrightarrow{S}}  \label{LL}
\end{equation}
where we put $\hbar=1$. Here $E$ is the energy as a functional of
the vector field $\overrightarrow{S}(x,t)$. Using the continuum
approximation (\ref{dSdx}), Hamiltonian (\ref{HH}) goes over into
the well-known energy functional:
\begin{equation}
E = \int\mathrm{d}x\left[ \frac{1}{2}
\left(\frac{\partial\overrightarrow{S}}{\partial x}\right)^2
+\alpha\left(s^2-S_z^2\right)\right] \label{EXXZ}
\end{equation}

Classical equation of motion (\ref{LL}) for model (\ref{EXXZ}) has
two constants of the motion: the magnetization (continuum analog to
a number of magnons)
\begin{equation}
M = s\int\mathrm{d}x(1-\cos\theta)  \label{M}
\end{equation}
where we subtract a constant to make $M$ finite for solitons, and
the momentum
\begin{equation}
P = s\int\mathrm{d}x(1-\cos\theta)\frac{\partial\varphi}{\partial x}
\label{P}
\end{equation}

The ground state of (\ref{EXXZ}) is the ferromagnetic configuration
with all spins parallel (or antiparallel) to the Z axis, i.e.
$\theta=0$ ($\theta=\pi$). The soliton configurations are specified
by the boundary conditions $\theta\to 0$ ($\theta\to\pi)$ at
$x\to\pm\infty$.

Fortunately, the classical equations of motion (\ref{LL}) for model
(\ref{EXXZ}) are exactly solvable \cite{Laksh}. In particular, the
energy for a given values of $M$ and $P$ is
\begin{equation}
E_{M,P}(\alpha) = 4s^2\sqrt{2\alpha}\frac{\cosh(M\sqrt{2\alpha}/2s)
-\cos(P/2s)}{\sinh(M\sqrt{2\alpha}/2s)}  \label{EMP}
\end{equation}

Remarkably, Eq.(\ref{EMP}) reproduces the exact result for the
energy at $\alpha\ll 1$ of the $M$-magnon bound state
\cite{Ovchinnikov} for the most quantum case $s=1/2$.

Here we will show that simple scaling arguments are able to
establish the correct scaling dependence for the energy. For this
aim we rescale the coordinate $x=\xi/\alpha^{1/2}$ and introduce the
normalized spin vector field $\overrightarrow{n}(x,t) =
\overrightarrow{S}(x,t)/s$ in Eq.(\ref{EXXZ}), which results in
\begin{equation}
E=s^2\alpha^{1/2}\int \mathrm{d}\xi \left[ \frac{1}{2}\left(
\frac{\partial\overrightarrow{n}}{\partial\xi}\right)^2 +(1-n_z^2)
\right] \label{EEE}
\end{equation}

We notice that the integrand in equation (\ref{EEE}) does not
depend on any parameters and is expressed through the normalized
spin vector field. This means that the energy scales as $E\sim
s^2\alpha^{1/2}$, which is correct for a kink or for a large
soliton excitations (see Eq.(\ref{EMP})). However, we can make one
more step. The same procedure for magnetization (\ref{M}) gives
\begin{equation}
M=s\alpha^{-1/2}\int\mathrm{d}\xi (1-n_z)
\end{equation}

This expression means that the magnetization forms a scaling
parameter $M\alpha^{1/2}/s$. In a similar manner one can find that
momentum (\ref{P}) produces a dimensionless parameter $P/s$
independent of $\alpha$.

Thus, the energy of a soliton of the size $M$ having a momentum $P$
can be written in a form:
\begin{equation}
E=s^2\alpha^{1/2}f(M\alpha^{1/2}/s,P/s)  \label{EscXXZ}
\end{equation}
where $f$ is a scaling function, which can not be found in the
framework of this scaling estimate. Fortunately, for model
(\ref{EXXZ}) this function is known exactly (\ref{EMP}), which
validates the above scaling arguments. Thus, simple scaling
estimates allowed us to establish the scaling parameters and the
scaling form for the soliton energy.

There is one more important fact which is worth noting here. The
exact static solution for a kink in the discrete XXZ model with
classical spins has been constructed by Gochev in Ref.\cite{gochev}.
He found that the kink energy has a form
\begin{equation}
E_{\mathrm{kink}} = 2s^2\sqrt{\Delta^2-1}  \label{Ekink}
\end{equation}
which reproduces the exact result for the case $s=1/2$
\cite{Ovchinnikov}. This allows us to assume that the energy of kink
(or large soliton, which has double energy of kink) for XXZ chain
with any value of $s$ has an universal dependence on $\Delta$
(\ref{Ekink}). We will see that this universality is destroyed in
case $\lambda\neq 0$.

\section{Classical continuum spin model near the IT point}

The NNN term in model (\ref{HH}) causes the frustration in the
system and immediately destroys the integrability of the model.
However, the behavior of the system for $0<\lambda<1/4$ and weak
easy-axis anisotropy remains very similar to the case $\lambda=0$.
In the classical and continuum approximation the Hamiltonian
reduces to an energy functional (\ref{EXXZ}) with renormalized
factor $(1-4\lambda)/2$ at $(\partial\overrightarrow{S}/\partial
x)^2$. In particular, this means that the found scaling form for
the soliton energy (\ref{EscXXZ}) remains valid.

The situation drastically changes near the IT point ($\lambda=1/4,\
\Delta=1$) where the ground state phase transition in the isotropic
case takes place. Let us deduce the energy functional describing the
vicinity of this point. The continuum approximation assumes that the
differences $(\theta_{n+1}-\theta_n)$ and
$(\varphi_{n+1}-\varphi_n)$ are small and Eq.(\ref{Edis}) can be
expanded in these differences. However, in the vicinity of the IT
point it is more clear and instructive to derive the continuum
approach in terms of a spin vector field. For this aim we rewrite
Hamiltonian (\ref{HH}) in the form:
\begin{equation}
H=\frac{1}{8}\sum (\mathbf{S}_{n+1}-2\mathbf{S}_n
+\mathbf{S}_{n-1})^2- \frac{\gamma}{8}\sum
(\mathbf{S}_{n+1}-\mathbf{S}_{n-1})^2 +\alpha\sum(s^2-S_n^z
S_{n+1}^z)  \label{H14}
\end{equation}
where the parameters $\alpha=\Delta-1$ and $\gamma=4\lambda-1$ are
small in the vicinity of the IT point.

In the classical and the continuum approach the expression
containing spin operators in the first term in Eq.(\ref{H14}) is
replaced by
\begin{equation}
\overrightarrow{S}_{n+1}-2\overrightarrow{S}_{n}+\overrightarrow{S}
_{n-1}\approx \frac{\partial^2\overrightarrow{S}(x_{n})}{\partial
x^2} \label{dSdx2}
\end{equation}
and that in the second term in Eq.(\ref{H14}) according to Eq.(\ref{dSdx}).

Thus, Hamiltonian (\ref{H14}) is mapped to the energy functional
\begin{equation}
E=\int \mathrm{d}x\left[ \frac{1}{8}\left( \frac{\partial^2
\overrightarrow{S}}{\partial x^{2}}\right)^2-\frac{\gamma}{2}\left(
\frac{\partial \overrightarrow{S}}{\partial x}\right) ^{2}+\alpha
\left(s^2-S_z^2\right) \right]  \label{Etr}
\end{equation}

The effect of the fourth-order term
$(\partial^2\overrightarrow{S}/\partial x^2)^2$ in the energy
functional like (\ref{Etr}) was studied before \cite{ivanov}, but it
was considered as a small correction to the main contribution given
by the term $(\partial\overrightarrow{S}/\partial x)^2$. On the
contrary for our model (\ref{Etr}) near the IT point the
fourth-order term $(\partial^2\overrightarrow{S}/\partial x^2)^2$
becomes the leading one. The appearance of the fourth-order term is
related to the fact that the one-magnon spectrum in the IT point
becomes $\varepsilon(k)\sim k^4$.

The energy functional (\ref{Etr}) in terms of the angular variables
$\varphi(x)$ and $\theta(x)$ has a rather cumbersome form:
\begin{eqnarray}
E &=&\frac{s^2}{8}\int \mathrm{d}x\left[ \theta^{\prime\prime
2}+\theta^{\prime 4}-4\gamma\theta^{\prime 2}+(\varphi^{\prime
\prime 2}+\varphi^{\prime 4}-4\gamma\varphi^{\prime 2}+8\alpha
)\sin^2\theta \right]
\nonumber \\
&&+\frac{s^{2}}{8}\int \mathrm{d}x\left[ (4-2\sin ^{2}\theta )\varphi
^{\prime 2}\theta ^{\prime 2}+(2\varphi ^{\prime \prime }\varphi ^{\prime
}\theta ^{\prime }-\varphi ^{\prime 2}\theta ^{\prime \prime })\sin (2\theta
)\right]  \label{Elong}
\end{eqnarray}
where the prime denotes the space derivatives $\partial/\partial x$.
One can check that Eq.(\ref{Elong}) represents the leading terms in
the expansion of Eq.(\ref{Edis}) in small differences
$(\theta_{n+1}-\theta_n)$ and $(\varphi_{n+1}-\varphi_n)$, which is
actually assumed in the continuum approximation.

At first, let us study the phase diagram of the classical continuum
model (\ref{Elong}). Numerical calculations (details will be given
below) confirmed a natural assumption that the ground state spin
configuration for the easy-axis anisotropy case ($\alpha>0$)
corresponds to the choice $\varphi=\mathrm{const}$. In this case the
energy functional (\ref{Elong}) simplifies to
\begin{equation}
E=\frac{s^2}{8}\int\mathrm{d}x(\theta^{\prime\prime 2}
+\theta^{\prime 4}-4\gamma\theta^{\prime 2}+8\alpha\sin^2\theta )
\label{Etheta}
\end{equation}

Variation of the energy functional (\ref{Etheta}) in $\theta(x)$
leads to the Euler equation
\begin{equation}
\frac{1}{4}\theta ^{\prime \prime \prime \prime }-\frac{3}{2}\theta ^{\prime
\prime }\theta ^{\prime 2}+\gamma \theta ^{\prime \prime }+\alpha \sin
(2\theta )=0  \label{eiler}
\end{equation}

The ground state in the ferromagnetic phase has a trivial solution
$\theta=0$ (or $\theta=\pi$) with zero energy. In the isotropic
case ($\alpha=0$) the transition from the ferromagnetic to the
spiral phase takes place at $\lambda=1/4$. For $\lambda>1/4$ the
solution $\theta(x)$ has a pure spiral form
\begin{equation}
\theta _{sp}(x)=\pm \sqrt{2\gamma }x  \label{sp}
\end{equation}
which evidently has zero mean spin projection on any axis
$\left\langle S_{x,y,z}\right\rangle=0$ and describes the spiral in
the XZ plane.

In the anisotropic case the spiral function (\ref{sp}) does not
satisfy Eq.(\ref{eiler}). However, it represents a good
approximation for the solution in the spiral phase. Using this
function as a variational one for the energy functional
(\ref{Etheta}) we find that the transition between the ferromagnetic
and the spiral phases occurs on the line $\gamma=\sqrt{\alpha}$.

Assuming that the correction to the spiral solution $\theta_{sp}(x)$
is small, we found from Eq.(\ref{eiler}) that the first correction
has the oscillating form:
\begin{equation}
\theta (x)\approx \sqrt{2\gamma }x-\frac{\alpha }{32\gamma^2}\sin
(\sqrt{8\gamma}x)  \label{spiral}
\end{equation}

This function is not a pure spiral: due to the oscillating
correction the spin vectors prefer to direct along the Z axis, so
that $\left\langle S_z^2\right\rangle >\left\langle
S_x^2+S_y^2\right\rangle $. However the mean values of all total
projections remain zero: $\left\langle S_{x,y,z}\right\rangle =0$.

The calculation of the energy functional (\ref{Etheta}) with the
function (\ref{spiral}) yields
\begin{equation}
\frac{E_{sp}}{s^2N}\approx
-\frac{\gamma^2-\alpha}{2}-\frac{\alpha^2}{128\gamma^2}
\label{Espiral}
\end{equation}

\begin{figure}[tbp]
\includegraphics[width=3in]{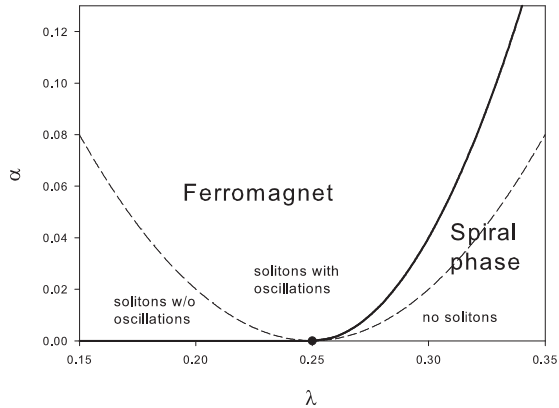}
\caption{The phase diagram of the classical F-AF model. Thick line
is the phase boundary between the ferromagnetic and the spiral
phases. Dashed lines are boundaries between regions of different
asymptotic regimes for large static solitons.} \label{phase}
\end{figure}

One can see that the correction to the energy is really small in the
spiral phase region. Energy (\ref{Espiral}) becomes negative for
$\gamma\gtrsim 0.992\sqrt{\alpha}$. This means that taking the
correction in Eq.(\ref{spiral}) into account shifts the transition
line from the ferromagnetic to the spiral phase on the value less
than $1\%$ (the smallness of the correction in Eq.(\ref{spiral})
allows us to leave the name spiral for this phase). The phase
transition from the ferromagnetic to the spiral phase at
$\gamma\approx 0.992\sqrt{\alpha}$ is obviously the first order one.
The transition line $\gamma=0.992\sqrt{\alpha}$ is shown in
Fig.\ref{phase}.

It is worth noting here that the found behavior of the transition
line corresponds to the classical version of spin model (\ref{HH})
and it is shifted for the quantum spin-$s$ model. Quantum
corrections for $s\geq 1$ change numerical coefficient in
Eq.(\ref{Espiral}). But for the $s=1/2$ case they lead to another
form of the transition line: $\gamma\sim\alpha^{3/5}$ \cite{DK08}.

Now let us study the excitations of model (\ref{Etr}). Similar to
the case $\lambda=0$ the lowest configuration in the ferromagnetic
region for a given $\left\langle S_z\right\rangle $ is described by
the static soliton-like solutions of Eq.(\ref{eiler}) with
$\varphi=\mathrm{const}$ over the ferromagnetic configuration
$\theta=0$ (or $\theta=\pi$). Unfortunately, we could not find the
exact soliton solution near the IT point. However, it is possible to
determine the scaling dependence of the energy of soliton of size
$M$ on parameters $\alpha$ and $\gamma$ using scaling estimates as
we did for the case $\lambda=0$. Near the IT point we perform the
rescaling $x=\xi/\alpha^{1/4}$ and
$\overrightarrow{S}(x,t)=s\overrightarrow{n}(x,t)$ which transforms
the energy functional (\ref{Etr}) to
\begin{equation}
E=s^2\alpha^{3/4}\int \mathrm{d}\xi \left[\frac{1}{8}\left(
\frac{\partial^2\overrightarrow{n}}{\partial
\xi^2}\right)^2-\frac{\mu}{2}\left(\frac{\partial
\overrightarrow{n}}{\partial\xi}\right)^2+\left( 1-n_z^2\right)
\right]  \label{Etrscal}
\end{equation}
with $\mu=\gamma/\sqrt{\alpha}$.

Since the integrand in Eq.(\ref{Etrscal}) depends on the parameter
$\mu$ only, we conclude that the energy of a kink or of a large
soliton is $E=s^2\alpha^{3/4}f(\mu)$. As a result of rescaling the
magnetization (\ref{M}) becomes:
\begin{equation}
M=s\alpha^{-1/4}\int\mathrm{d}\xi (1-n_z)  \label{Mtr}
\end{equation}

So, the magnetization produces a scaling parameter
$\nu=\alpha^{1/4}M/s$. The momentum (\ref{P}) forms the same
parameter $k=P/s$ as for the case $\lambda=0$.

Thus, the energy of a soliton of the size $M$ can be represented in
a form:
\begin{equation}
E = s^2\alpha^{3/4}f(\mu,\nu,k)  \label{Escal}
\end{equation}
where the scaling function $f$ can be found numerically. This
scaling form is exactly the same as was found for the quantum
spin-$1/2$ model near the IT point \cite{DK09}.

We note that similar scaling procedure for the $D$-dimensional
version of model (\ref{Etr}) gives the scaling form for the energy
of the soliton of the size $M$ (even small-amplitude solitons are
stable for model (\ref{Etr}) in two- and three-dimensional cases
\cite{Kosevich}):
\begin{eqnarray}
E &=&s^2\alpha^{1/2}f(\mu,\alpha^{1/2}M/s,k),\qquad D=2  \nonumber \\
E &=&s^2\alpha^{1/4}f(\mu,\alpha^{3/4}M/s,k),\qquad D=3
\end{eqnarray}
and the case $D=4$ requires special treatment.

Though the exact solution of the corresponding equation of motion
(\ref{LL}) is unknown it is possible to identify the necessary
conditions for a stability of solitons and establish their
asymptotic behavior. It is convenient to introduce the complex
function
\begin{equation}
\psi = n_x + i n_y  \label{psi}
\end{equation}
so that $\psi=e^{i\varphi}\sin\theta$ and $n_z^2=1-|\psi|^2$.

Equation of motion (\ref{LL}) for $\psi(\xi,\tau)$ near the IT
point reads
\begin{equation}
i\frac{\partial\psi}{\partial\tau}=
\frac{1}{4}n_z\frac{\partial^4\psi}{\partial\xi^4}
-\frac{1}{4}\psi\frac{\partial^4n_z}{\partial\xi^4} +\mu
n_z\frac{\partial^2\psi}{\partial\xi^2}
-\mu\psi\frac{\partial^2n_z}{\partial\xi^2} +2n_z\psi \label{LLpsi}
\end{equation}
where $\tau=s\alpha^{3/4}t$ is rescaled time.

Solitons are localized objects and, therefore, it requires that
$|\psi|\to 0$ far from the soliton center. So, at large distance
from the center one can linearize Eq.(\ref{LLpsi}) in $\psi$ by
putting $n_z=1$:
\begin{equation}
i\frac{\partial\psi}{\partial\tau}
=\frac{1}{4}\frac{\partial^4\psi}{\partial\xi^4}+\mu
\frac{\partial^2\psi}{\partial\xi^2}+2\psi \label{psilinear}
\end{equation}

We seek the asymptotic of $\psi$ in conventional exponential form
\begin{equation}
\psi(\xi,\tau)=\exp\left[-i\omega\tau-\kappa(\xi-v\tau)\right]
\label{psiasymp}
\end{equation}
where $v$ and $\omega$ are normalized linear and angular
velocities defined as
\begin{eqnarray}
\omega  &=&\frac{1}{\alpha s}\frac{\partial E}{\partial M}
=\frac{\partial f(\mu,\nu,k)}{\partial\nu} \nonumber \\
v &=&\frac{1}{\alpha^{3/4}s}\frac{\partial f(\mu,\nu,k)}{\partial k}
\label{omegav}
\end{eqnarray}

Substituting Eq.(\ref{psiasymp}) into Eq.(\ref{psilinear}) we
obtain equation for $\kappa$
\begin{equation}
\frac{\kappa^4}{4}+\mu\kappa^2-iv\kappa +2-\omega =0 \label{eqkappa}
\end{equation}

This equation has four roots. The condition of the soliton
existence is that all roots have non-zero real part, which secures
the decay of solutions at $|\xi-v\tau|\to\infty$. The region in
$(v,\omega)$ plane where $\psi$ exponentially vanishes at
$|\xi-v\tau|\to\infty$ is defined as $\omega<\omega_0(v)$. The
dependence $\omega_0(v)$ is parametrically definable function
\begin{eqnarray}
v &=& q^3-2\mu q  \nonumber \\
\omega_0 &=& 2 +\mu q^2 -\frac34q^4 \label{stable}
\end{eqnarray}
where the parameter $q$ runs from $-\infty$ to $\infty$ for $\mu<0$
and $|q|\geq\sqrt{2\mu}$ for $\mu>0$.

\begin{figure}[tbp]
\includegraphics[width=3in]{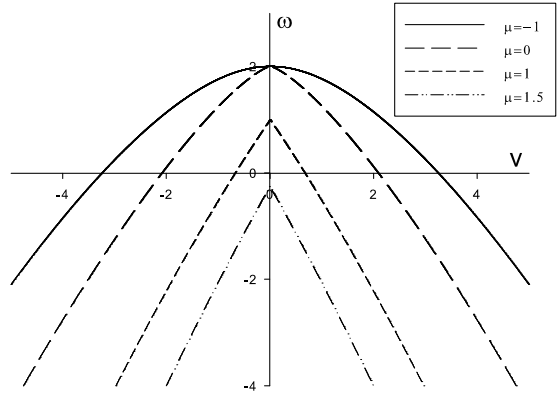}
\caption{Boundaries of the existence regions for soliton in
$(v,\omega)$ plane for different $\mu$. Allowable values of $v$ and
$\omega$ lie below the boundaries.} \label{vomega}
\end{figure}

Equations (\ref{stable}) of the boundary of the soliton existence
region coincide with the condition of the spin-wave instability.
The existence regions for a few values of the parameter $\mu$ are
shown in Fig.\ref{vomega}. For $\mu<0$ the region has a form
similar to the case $\lambda=0$ with the quadratic dependence
$\omega_0(v)$ near the maximum point ($v=0,\ \omega=2$). For the
case $\mu=0$ this dependence becomes $\omega_0=2-3v^{4/3}/4$. When
$\mu>0$, the dependence $\omega_0(v)$ shifts down and a cusp
appears at the maximal point ($v=0,\ \omega=2-\mu^2$).

The region of the soliton existence contains allowable values of
$v$ and $\omega$ for a given $\mu$. As it can be seen in
Fig.\ref{vomega} the soliton can exist in the spiral phase
($\mu>0.992$). But, for example, the kink is unstable for
$\mu>\sqrt{2}$ since the point ($v=0,\ \omega=0$) corresponding to
the kink does not belong to the existence region. It is worth
noting that one should be careful in treating of the part of the
existence region with large negative $\omega$, where the
derivatives of $\theta(x)$ and $\varphi(x)$ near the center of
soliton become large and the continuum approximations
Eqs.(\ref{dSdx}), (\ref{dSdx2}) become doubtful.

Within the existence region the roots of Eq.(\ref{eqkappa}) have
the forms
\begin{eqnarray}
\kappa_{1,3} &=&\pm a_1 + i b  \nonumber \\
\kappa_{2,4} &=&\pm a_2 - i b \label{k1234}
\end{eqnarray}
where $a_1$, $a_2$ and $b$ are real nonnegative quantities and
$a_1\leq a_2$. Certainly, one should take into account only the
roots providing the decay of $|\psi|$ at the corresponding limit.
For determinacy we will consider the limit $(\xi-v\tau)\to\infty$
and, therefore, analyze the roots $\kappa_1=a_1+ib$ and
$\kappa_2=a_2-ib$. At infinity only the root $\kappa_1$ with the
smaller real part is vital. So, the asymptotes of the angles at
$(\xi-v\tau)\to\infty$ are $\theta\sim\exp[-a_1(\xi-v\tau)]$ and
$\varphi=-\omega\tau-b(\xi-v\tau)$.

The case of the static soliton ($v=0$) requires special treatment,
because in this case both roots $\kappa_1$ and $\kappa_2$ have
equal real parts $a_1=a_2$ (they immediately split if $v\neq0$).
Fortunately, for the static soliton the solution of
Eq.(\ref{eqkappa}) can be obtained explicitly and this allows to
analyze in detail the behavior of the soliton asymptotes for given
values of $\mu$ and $\omega$. The decaying roots at
$(\xi-v\tau)\to\infty$ are
\begin{equation}
\kappa_1 =\sqrt{-2\mu -2\sqrt{\mu^2-2+\omega}} \label{kappa}
\end{equation}
and $\kappa_2=\kappa_1^*$.

Simple analysis of Eq.(\ref{kappa}) shows that there are three
regimes of the asymptotic (\ref{psiasymp}). The exponent
$\kappa_1$ is real for $\mu<-\sqrt{2-\omega }$, which means that
$\theta(\xi)$ smoothly tends to zero at infinity. This region lies
totally in the ferromagnetic phase. In the region $-\sqrt{2-\omega
}<\mu<\sqrt{2-\omega}$ the parameter $\kappa_1$ contains both real
and imaginary parts. Here the superposition of asymptotes
(\ref{psiasymp}) with $\kappa_1$ and $\kappa_2=\kappa_1^*$ results
in the decay with oscillation at infinity: $\theta(\xi)\sim
e^{-a_1\xi}\sin(b\xi)$ and $\varphi=\rm{const}$. Finally, for
$\mu>\sqrt{2-\omega}$ the parameter $\kappa_1$ is pure imaginary,
which implies that the solution oscillates but does not decay at
infinity. Therefore, in this region there are no static
soliton-like excitations.

As was noted above only one asymptotic with $\kappa_1$ having
smaller real part survives at infinity for $v\neq 0$. This implies
that the asymptotic of the function $\theta(\xi)$ has no zeros.
However, in the region close to the soliton center the function
$\theta(\xi)$ can have a finite number of zeros and it is really
observed numerically.

Generally, the relation between the parameters $\omega,v$ and
$\nu,k$ is unknown analytically. However, it can be found for
large values of $\nu$. For large solitons ($\nu\gg 1$) the soliton
energy and, therefore, the scaling function $f(\mu,\nu,k)$
saturates to some finite value. This means that for $\nu\gg 1$
both $\omega\to 0$ and $v\to 0$. The boundaries between the
regions corresponding to different asymptotic regimes for large
solitons are obtained from the above analysis and they are shown
in Fig.\ref{phase}. As one can see in Fig.\ref{phase} the
oscillating large solitons exist not only in the ferromagnetic
phase but, partly, in the spiral phase region
$0.992<\mu<\sqrt{2}$. The amplitude of the soliton oscillations
grow as $\mu$ increases and the oscillations are most strong in
the spiral phase. However, one should remember that the soliton
excitations in the spiral phase lie in the high-energy part of the
spectrum, because the energy of the soliton-like solutions differs
from the ferromagnetic one on some finite value (\ref{Escal}).
Therefore, the soliton energy is higher than the spiral one
(\ref{Espiral}) on the value proportional to the system size.
Nevertheless, these states can be important in strong magnetic
field close to the saturation value.

If we expand $n_z\approx 1-|\psi|^2/2$ and take into account the
terms $\sim|\psi|^2\psi$ in Eq.(\ref{LLpsi}), then Eq.(\ref{LLpsi})
near the border of the soliton existence reduces to the modified
non-linear Schr\"{o}dinger equation containing the fourth-order
derivative term
\begin{equation}
i\frac{\partial\psi}{\partial\tau}
=\frac{1}{4}\frac{\partial^4\psi}{\partial\xi^4}
+\mu\frac{\partial^2\psi}{\partial\xi^2} +2\psi
-\frac{\omega+iv\kappa}{2}\left\vert\psi\right\vert^2\psi
\label{LLpsi2}
\end{equation}

Unfortunately, in spite of substantial simplification of the initial
equation of motion (\ref{LLpsi}), the exact solution of this
equation is unknown and we have to use numerical calculation.

We note that Eqs.(\ref{psiasymp}),(\ref{eqkappa}) are valid for
small $\alpha\ll 1$, but for any $\gamma$. In particular, for the
case $\lambda=0$ ($\gamma=-1$) they correctly reproduce the known
asymptotic for $\theta$ and $\varphi$.

\section{Numerical calculations}

We have carried out a numerical analysis of the discrete and the
continuum versions of the classical spin model (\ref{HH}) in the
vicinity of the IT point. Both energy functionals (\ref{Edis}) and
(\ref{Elong}) were minimized numerically over angles
$\theta_n,\varphi_n$ and $\theta(x),\varphi(x)$ with fixed values
of magnetization (\ref{M}) and momentum (\ref{P}). The periodic
boundary conditions for soliton and the open boundary conditions
for kink excitations were imposed. The numerical calculations
showed that for small $\alpha$ the difference between the discrete
and the continuum models is negligible as expected and it
increases as $\alpha$ grows. On the investigation of the dynamics
of solitons we restrict ourself to small anisotropy $\alpha\ll 1$,
when the continuum approximation is valid, because in the discrete
model with finite $\alpha$ a so-called pinning potential appears
\cite{Mikeska,IM} and the momentum becomes undefined. So, on
default we will present numerical results for the discrete
classical model (\ref{Edis}), keeping in mind that the continuum
approach corresponds to the limit $\alpha\to 0$.

We studied the finite size effects for both discrete and continuum
models and found that the system size $N$ should be taken so that
the parameter $\alpha^{1/4}N\gg 1$. The finite size effects will be
discussed in detail in Sec.V. Here we only notice that when the
relation $\alpha^{1/4}N\gg 1$ is fulfilled the convergence of a
solution accelerates exponentially with $N$.

At first we present and discuss the results for small values of
$\alpha$ and $\gamma$, when the continuum approximation is justified
and the soliton energy takes the scaling form (\ref{Escal}). In
order to verify the found scaling equation we plotted the results
for a fixed $k$ and different values of $\alpha$, $\gamma$, $M$ as
$E/(s^2\alpha ^{3/4})$ vs. scaling parameters $\nu=\alpha^{1/4}M/s$
and $\mu=\gamma/\sqrt{\alpha}$. For small $\alpha$ and $\gamma$ all
data must lie on one curve which actually represents the scaling
function $f(\mu,\nu,k)$.

\begin{figure}[tbp]
\includegraphics[width=3in]{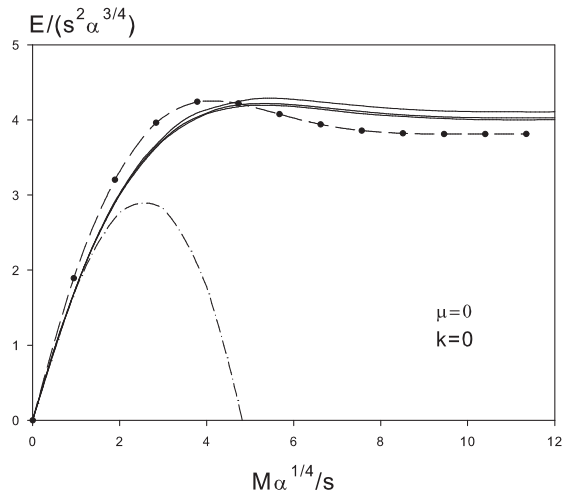}
\caption{Scaled energy $E/s^2\alpha^{3/4}$ versus scaled soliton
size $\nu=M\alpha^{1/4}/s$ for $\mu=0$ and $k=0$. Solid lines are
soliton energies for three values of $\alpha=10^{-2},\ 10^{-3},\
10^{-4}$ from top to bottom. Circles (joined by dashed line) are
energies of magnon complexes of quantum $s=1/2$ F-AF chain for
$N=24$ and $\alpha=0.05$ \cite{DK09}. Dot-dashed line is the energy
of boson bound states of model (\ref{Hbose2}).} \label{emmu0}
\end{figure}

For example, the numerical data for the static solitons ($k=0$)
and for $\mu=0$ is shown in Fig.\ref{emmu0}. Here the calculated
dependencies of the soliton energy on the soliton size $M$ for
three values of $\alpha =10^{-2}$, $10^{-3}$, $10^{-4}$ are
demonstrated in axes $E/(s^2\alpha^{3/4})$ vs.
$\nu=\alpha^{1/4}M/s$. As we see all three solid curves lie very
close to each other and they rapidly converges in the limit
$\alpha\to 0$, so that the curve corresponding to $\alpha=10^{-4}$
perfectly describes the scaling function $f(0,\nu,0)$.

The function $f(0,\nu,0)$, shown in Fig.\ref{emmu0}, reaches a
maximum at $\nu\approx 5$ and then rapidly saturates, yielding the
energy of large solitons $E\approx 4s^2\alpha^{3/4}$. For comparison
we put also in Fig.\ref{emmu0} similar data for $k=0$ and
$\lambda=1/4$ obtained in Ref.\cite{DK09} for the quantum spin-$1/2$
chain of length $N=24$ and $\alpha=0.05$, which represents a rough
estimate for the spin-$1/2$ scaling function $f_{1/2}(0,\nu,0)$. One
can see that the behavior of the scaling functions for the quantum
$s=1/2$ and the classical $s\to\infty$ models are very similar,
though they have different limits at $\nu\to\infty$. Thus, we
believe that the scaling law Eq.(\ref{Escal}) is valid for quantum
model (\ref{HH}) with general $s$ and the corresponding scaling
functions $f_s$ behave similar to $f_{1/2}$ and $f$.

\begin{figure}[tbp]
\includegraphics[width=3in]{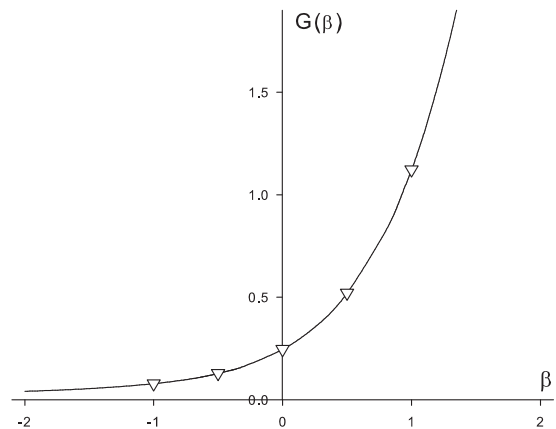}
\caption{The functions $G(\beta)$ for the energy of small static
solitons (triangles) and $G_b(\beta)$ for the energy of boson
complexes Eq.(\ref{Eboson}) (solid line).} \label{Gbeta}
\end{figure}

It is known \cite{Schneider} that for $\lambda=0$ there is an
equivalence between the energies of the classic static solitons of
small size and the bound boson complexes of the Bose-Hamiltonian
which is mapped from the anisotropic Heisenberg model. We expect
that this mapping is valid for $\lambda\neq 0$ as well and it
allows to determine the functional form of the energy for small
solitons near the IT point. As it will be shown in Sec.VI the
energy of the $M$-boson bound state of the corresponding
Bose-model is
\begin{equation}
\frac{E}{s^2\alpha^{3/4}}=2\nu-\nu^{7/3}G_b(\beta) \label{EGB}
\end{equation}
where $\beta=\mu/\nu^{2/3}$ and the dependence $G_b(\beta)$ is shown
in Fig.\ref{Gbeta}.

We found that the numerical data for $f(\mu,\nu,0)$ at $\nu\ll 1$
is perfectly fitted by Eq.(\ref{EGB}) and the function $G(\beta)$
for the classical solitons perfectly coincides with $G_b(\beta)$
(see Fig.\ref{Gbeta}). Therefore, we believe that the scaling
functions for solitons and boson complexes are the same at $\nu\ll
1$ and are given by Eq.(\ref{EGB}). However, they are different
generally (see Fig.\ref{emmu0}). In particular, at $\nu\gg 1$ the
energy of the soliton saturates while the boson complex energy
diverges.

For small solitons with $k=0$ according to Eq.(\ref{omegav})
$\omega$ is given by the derivative of the right-hand side of
Eq.(\ref{EGB}). Therefore, the boundary between different regimes of
asymptotes given by equation $\mu^2=2-\omega$ (see Eq.(\ref{kappa}))
transforms for small solitons to the relation $3\beta^2=7G-2\beta
G'$ for $\beta$. Numerical calculation of the function $G(\beta)$
(Fig.\ref{Gbeta}) indicates that there is only one solution
$\beta=-0.57$ of the above relation. This means that the oscillation
behavior of the solitons of size $M$ exists if $\mu>-0.57\nu^{2/3}$
and that the first two terms in the expansion in small $\nu$
(\ref{EGB}) is not sufficient to determine the boundary where
solitons disappear.

\begin{figure}[tbp]
\includegraphics[width=3in]{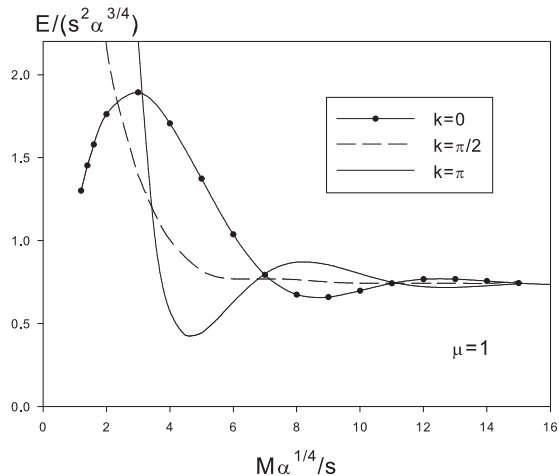}
\caption{Scaled energy $E/s^2\alpha^{3/4}$ versus scaled soliton
size $\nu=M\alpha^{1/4}/s$ for $\mu=1$ and for different $k$.}
\label{emmu1}
\end{figure}

The scaling function $f(\mu,\nu,k)$ for $k=0,\ \pi/2,\ \pi$ at
$\mu=1$ is demonstrated in Fig.\ref{emmu1}. As can be seen in
Fig.\ref{emmu1} the scaling function $f(1,\nu,k)$ converges at
$\nu\to\infty$ to the same finite value for all $k$, which is
natural, because the spectrum of large solitons is flat. Another
observation followed from Fig.\ref{emmu1} is that the scaling
function for $\mu=1$ oscillates substantially stronger than that in
the case $\mu=0$ (Fig.\ref{emmu0}).

As we have stated above the necessary condition for the soliton
existence is
\begin{equation}
\frac{\partial f(\mu,\nu,k)}{\partial\nu} \leq \omega_0(\mu,v)
\label{stab}
\end{equation}
which means that $\partial f/\partial\nu\leq 2$ for $\mu\leq 0$
and $\partial f/\partial\nu\leq 2-\mu^2$ for $\mu\geq 0$ for any
velocity.

This inequality imposes certain limitations on the soliton size
(or the allowable values of $\nu$) with given values $\mu$ and
$k$. However, the existence regions in the ($\nu,k$) plane can not
be obtained directly from that in ($v,\omega$) plane, because the
analytical relations between $\nu,\ k$ and $v,\ \omega$ near the
IT point are unknown in contrast with the integrable case
$\lambda=0$. Therefore, the stability of the soliton for different
values of parameters $\mu$, $\nu$ and $k$ can be established by
numerical minimization of the energy functional only. Our
calculations show that for $\mu<0$ the solitons of any size and
for all $k$ exist and inequality (\ref{stab}) is satisfied.
However, the situation changes for $\mu>0$. For example, the
solitons with $\mu=1$ and $k=0$ are stable for $\nu\gtrsim 1$ only
as it is shown on Fig.\ref{emmu1}. As a matter of fact for $\mu>0$
only part of a whole phase space (a half-strip $\nu>0,\ 0<k<\pi$)
is allowable for solitons. In general, the allowable region in
($\nu,k$) half-strip has a complicated form and can consist of
many disconnected parts. For instance, at $\mu=1.5$ the static
solitons with $k=0$ are stable in the range $3.5<\nu<8$ though for
$k\neq 0$ small solitons exist.

Though a detailed phase picture can be find only numerically,
nevertheless we can make some statements about it. Since soliton
is a localized object, its energy inevitably saturates at
$\nu\to\infty$ to some finite value independent of $k$. This means
that for large solitons $\partial f/\partial\nu\to 0$. Therefore,
for $\mu<\sqrt{2}$ all large enough solitons with
$\nu>\nu_{\min}(\mu)$ are allowable. On the other hand deeply in
the spiral phase when $\mu>\sqrt{2}$ the allowable region is
restricted by $\nu<\nu_{\max}(\mu)$ and large solitons do not
exist. In fact, we did not find large solitons for $\mu>\sqrt{2}$
in our calculations.

It is interesting to study the constraints on the soliton size in
the isotropic limit of model (\ref{HH}) at $\lambda>1/4$, when
$\alpha\to 0$ and $\mu\to\infty$. It can be shown that the
function $f(\mu,\nu,k)$ at $\nu\ll 1$ and any $k\neq 0$ behaves as
($const/\nu^3$), where $const$ does not depend on $\alpha$. In
this case condition (\ref{stab}) reduces to $\gamma M^2<const$. It
means that the maximal possible size of soliton grows at
$\gamma\to 0$. This fact qualitatively agrees with the observed
behavior \cite{Kecke,Laeuchli} of the stability of magnon bound
complexes with $k=\pi$ in the isotropic quantum $s=1/2$ F-AF model
at $\lambda=1/4$. This is one more indication of the resemblance
of the classical solitons and the quantum bound complexes in the
F-AF model.

\begin{figure}[tbp]
\includegraphics[width=3in]{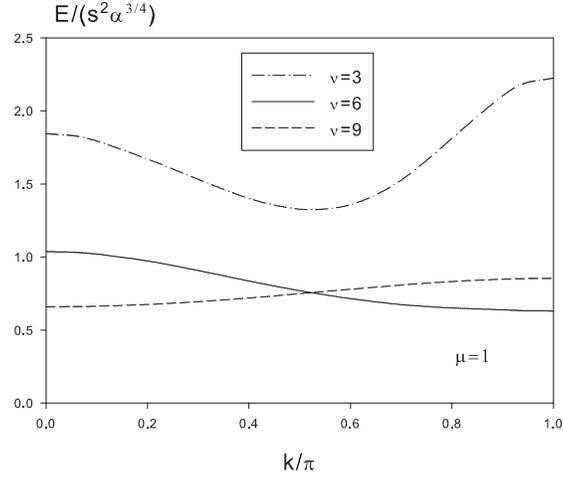}
\caption{Energy momentum spectra at $\mu=1$ for $\nu=3$ (dot-dashed
line), $\nu=6$ (dashed line) and $\nu=9$ (solid line).} \label{EP}
\end{figure}

In Fig.\ref{EP} the dependence of the soliton energy on the
momentum for $\mu=1$ is plotted for three different soliton sizes
corresponding to $\nu=3$, $\nu=6$ and $\nu=9$. It is interesting
to note that similarly to the case $\lambda=0$, there is a
periodical dependence of the energy on the momentum in the
\emph{continuum} model near the IT point. We see that the
dependence of the energy on the momentum rapidly flattens with the
increase of the soliton size. In other words, the soliton mass
exponentially grows with the soliton size. In this respect the
behavior of the soliton spectrum is similar to the exactly
solvable case $\lambda=0$ (see Eq.(\ref{EMP})). However, the
dependence $E(k)$ has a very complex form in contrast to the case
$\lambda=0$. As it is shown on Fig.\ref{EP} $k_{\min}$ can be
$k=0$, $k=\pi$ or even intermediate value between them. For
$\nu=6$ the minimum of $E(k)$ is obtained at $k=\pi$. With the
subsequent increase of the soliton size $\nu$ the form of the
spectrum alternates: it has the minimum either at $k=0$ or at
$k=\pi$. However, for $\nu>10$ the spectrum becomes so flat that
it is difficult to distinguish these cases numerically. The
soliton velocity $v$ is zero in both points $k=0$ and $k=\pi$.

\begin{figure}[tbp]
\includegraphics[width=3in]{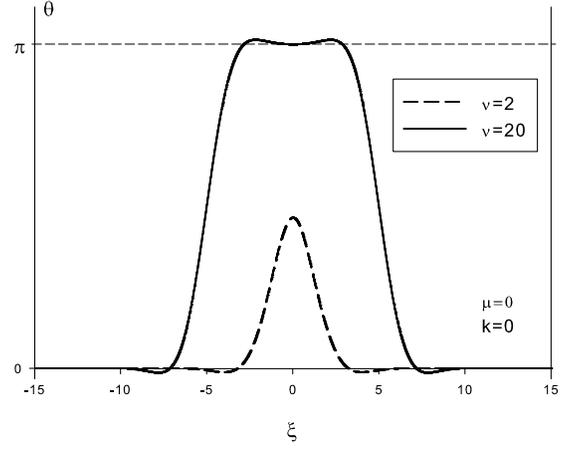}
\caption{Shapes of small ($\nu =2$) and large ($\nu=20$) solitons
for $\mu=0$ and $k=0$.} \label{solitonmu0p0}
\end{figure}

\begin{figure}[tbp]
\includegraphics[width=3in]{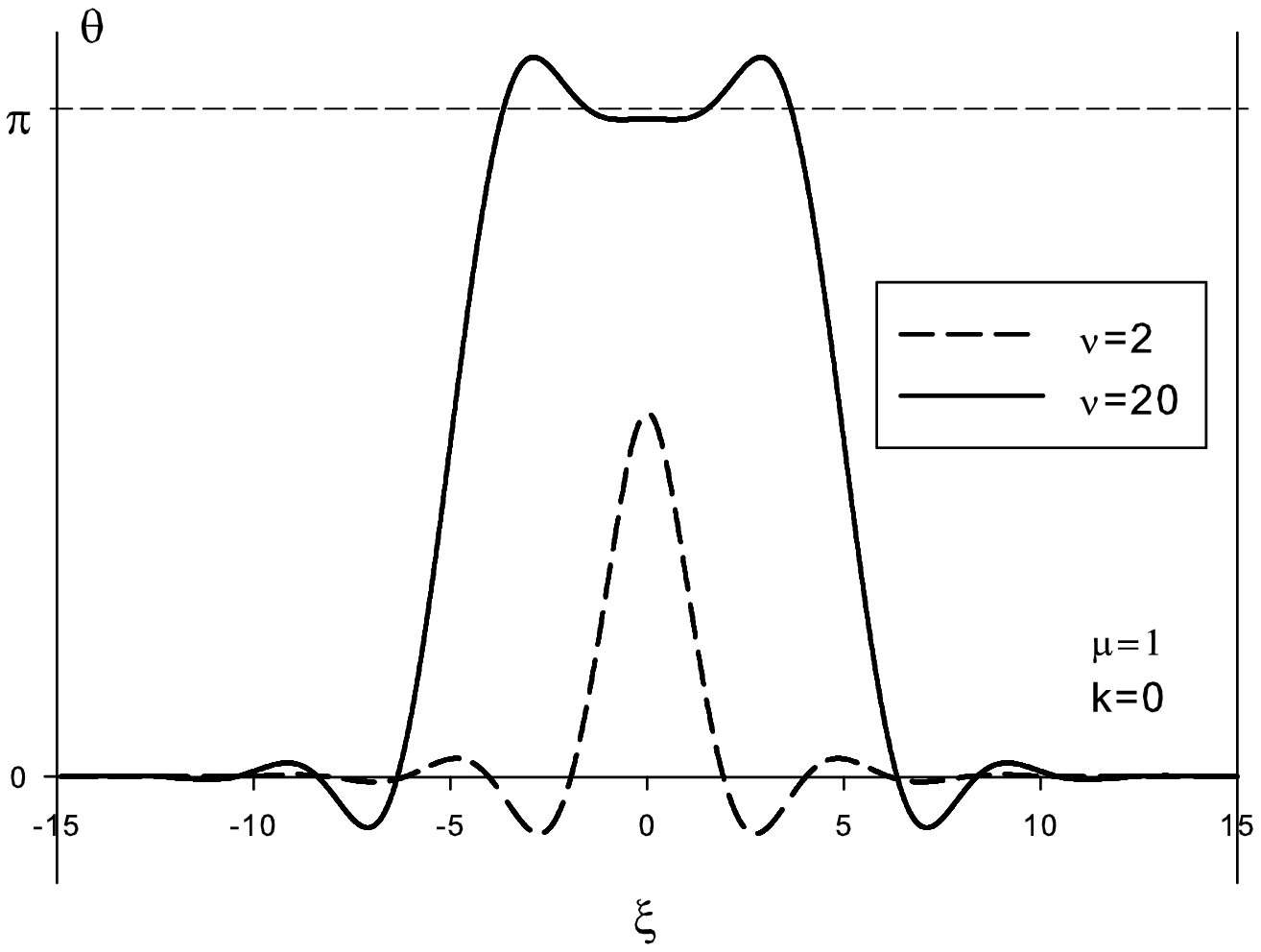}
\caption{Shapes of small ($\nu =2$) and large ($\nu=20$) solitons
for $\mu=1$ and $k=0$.} \label{solitonmu1p0}
\end{figure}

\begin{figure}[tbp]
\includegraphics[width=3in]{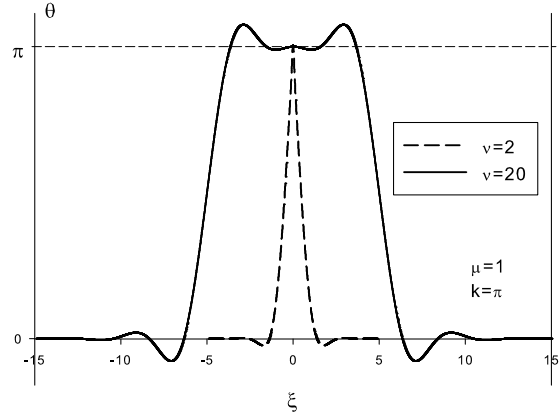}
\caption{Shapes of small ($\nu =2$) and large ($\nu=20$) solitons
for $\mu=1$ and $k=\pi$.} \label{solitonmu1ppi}
\end{figure}

We have obtained numerically the soliton shapes for different
values of the parameters $\mu$, $\nu$ and $k$. The shapes of
static ($k=0$) solitons at $\mu=0$ for the cases $\nu=2$ and
$\nu=20$ are demonstrated in Fig.\ref{solitonmu0p0}. The soliton
solutions oscillate around $\theta=0$ (and $\theta=\pi$ for large
soliton) and these oscillations exponentially decay as moving off
the domain walls. Numerical calculations show that at approaching
to the spiral phase the amplitude of the oscillations grows and
its damping decreases with the distance from the soliton center.
The shape of solitons of two different sizes with $\nu=2$ and
$\nu=20$ at $\mu=1$, which is very close to the transition line,
is shown in Fig.\ref{solitonmu1p0}. Here one can see the increase
of oscillations in comparison with the case $\mu=0$. The observed
behavior of the soliton shape is in full accord with conclusions
followed from the analysis of Eq.(\ref{kappa}).

The shape of solitons changes with the momentum $k$. This change
is more pronounced for small solitons: they become narrower and
higher. At the same time the shapes of large solitons remain
almost the same. As an example the shapes of small and large
solitons for the momentum $k=\pi$ and $\mu=1$ are demonstrated in
Fig.\ref{solitonmu1ppi}. As shown in Fig.\ref{solitonmu1ppi} the
function $\theta(x)$ has a cusp at the maximum, which resembles
the case $\lambda=0$. However, the behavior of the function
$\varphi(\xi)$ drastically differs from the case $\lambda=0$
\cite{KosJETP}. In particular, $\varphi(\xi)$ is discontinuous at
$\xi=0$ in contrast with that for $\lambda=0$, where this function
is linear near $\xi=0$.

One more illustration of the oscillation behavior of localized
excitations in the F-AF model is the kink solution for the
particular case $\mu=0$ ($\lambda=1/4$). Though we did not find
the analytic solution of the non-linear equation (\ref{eiler}), we
found for this particular case some signs that it can admit a
solution in a closed form. Namely, we studied numerically the kink
excitation described by equation (\ref{eiler}) with $\gamma=0$.
The boundary conditions are $\theta(-\infty)=0$,
$\theta(\infty)=\pi$ and
$\theta^{\prime}(-\infty)=\theta^{\prime}(\infty)=0$. Numerical
solution showed with very high accuracy that the asymptotic of the
kink solution centered at $\xi=0$ has a form
\begin{eqnarray}
\theta (\xi &\to &-\infty )=2\sin (1)\cos (-2^{1/4}\xi -1)e^{2^{1/4}\xi }
\nonumber \\
\theta (\xi &\to &\infty )=\pi -2\sin (1)\cos (2^{1/4}\xi -1)e^{-2^{1/4}\xi }
\end{eqnarray}
The oscillation behavior of the kink asymptotes is in full accord
with Eq.(\ref{kappa}).

At $|\xi|\ll 1$
\begin{equation}
\theta(\xi)=\frac{\pi}{2}+2^{1/4}\xi
\end{equation}

Besides, the calculated kink energy has a surprisingly simple form
\begin{equation}
E = 2s^2\alpha^{3/4}
\end{equation}
where the numerical factor $2$ was found with high precision.
Moreover, the contributions to this kink energy from different
terms in Eq.(\ref{Etheta}) have simple rational ratios. All these
facts together give a hope to obtain the exact solution of
Eq.(\ref{eiler}) for $\gamma=0$ in closed form.

\begin{figure}[tbp]
\includegraphics[width=3in]{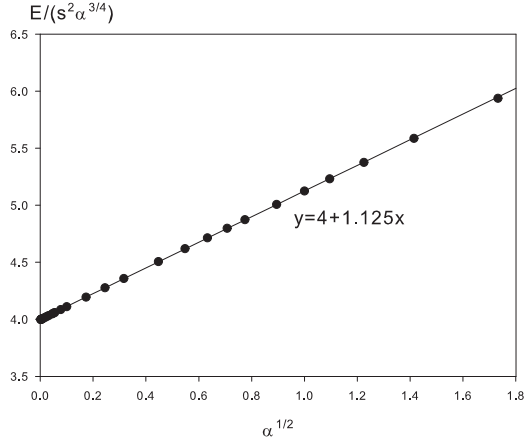}
\caption{Dependence of the scaled energy of large solitons on the
anisotropy $\sqrt{\alpha}$ for $\lambda =1/4$.} \label{Ealpha}
\end{figure}

The continuum approximation is valid for $\alpha\ll 1$. If the
parameter $\alpha$ is not very small there are corrections to the
soliton energy due to a discreteness. The energy of large soliton
($\nu\gg 1$) at $\lambda=1/4$ obtained numerically is shown in
Fig.\ref{Ealpha}. As one can see the energy is perfectly described
by equation
\begin{equation}
E/s^2 \approx 4\alpha^{3/4} +1.12\alpha^{5/4}
\end{equation}
which is valid up to $\alpha\sim 1$. Here the first term is
described by the continuum approximation, while the second term
represents the correction coming from the discreteness of the
lattice. The found correction to the soliton energy
($\sim\alpha^{5/4}$) differs from the correction to the energy of
multimagnon complex ($\sim\alpha$) found for the case $s=1/2$ in
Ref.\cite{DK09}.

We found that similar behavior of the energy of the large soliton
takes place for any $\mu$. So, the energy for $\nu\gg 1$ can be
written as
\begin{equation}
E = s^2\alpha^{3/4}f(\mu) + s^2\alpha^{5/4}g(\mu)  \label{E3454}
\end{equation}
where $f(\mu)$ is the saturated values of $f(\mu,\nu,k)$ at
$\nu\to\infty$.

\begin{figure}[tbp]
\includegraphics[width=3in]{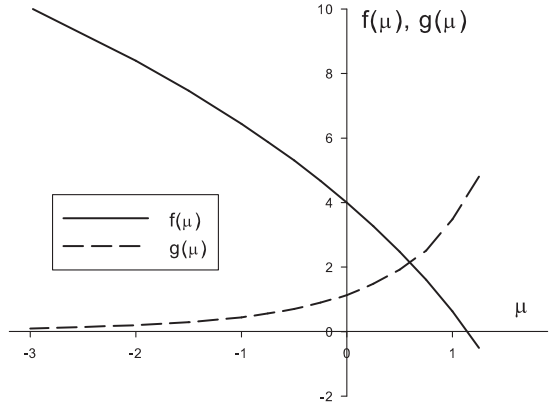}
\caption{Functions $f(\mu)$ (solid line) and $g(\mu)$ (dashed line)
describing two leading terms for the energy of large solitons
(Eq.(\ref{E3454})).} \label{fgmu}
\end{figure}

The behaviors of the functions $f(\mu)$ and $g(\mu)$ are shown in
Fig.\ref{fgmu}. As one can see $f(\mu)$ goes down with the
increase of $\mu$ and becomes negative in the spiral phase at
$\mu>1.2$. At the same time, the correction in Eq.(\ref{E3454})
increases at approaching to the spiral phase.

In the opposed limit of large anisotropy $\alpha\gg 1$, the
asymptotic of the soliton energy is found using a simple
perturbation theory, which gives
\begin{equation}
E/s^{2}=4\alpha -8\lambda -\frac{2}{\alpha}
-\frac{2\lambda^2}{\alpha s} + O(\alpha^{-2})
\end{equation}

As follows from this equation the energy is not simply proportional
to $s^2$ in contrast with the case $\lambda=0$ (Eq.(\ref{Ekink})).
This means that the dependence of the energy of kink or large
soliton is not a universal function of $\alpha$ for general $s$ as
it is for $\lambda=0$.

\section{Finite-size effects. Transition from uniform to localized solution.}

In the preceding sections we investigated the soliton and the kink
excitations of model (\ref{Edis}) in the infinite system. Now we
study the finite-size effects of these states. An interesting
property of the soliton solutions of the classical model
(\ref{Edis}) on finite ring is the existence of the critical value
$\alpha_0(L)$ below which the uniform solution
($\theta(x)=\mathrm{const}$) is realized. At $\alpha>\alpha_0$
this state develops into localized soliton. This transition from
the uniform to the localized solution is similar to the well-known
Gross-Pitaevskii transition in the Bose-systems \cite{BEGP}.

We consider the static soliton on the ring of size $L$ with
periodic boundary conditions. We use the continuous approach
because the finite-size effects are essential at small values of
$\alpha$. In the beginning we study the case $\lambda=0$. In this
case the function $\theta(x)$ providing the extremum of the energy
functional (\ref{EXXZ}) satisfies the Lagrange-Euler equation
\begin{equation}
\frac{\partial^2\theta}{\partial x^2}=\alpha\sin(2\theta)
-\alpha\omega\sin\theta ,\qquad -\frac{L}{2}<x<\frac{L}{2}
\label{eqXXZ}
\end{equation}
with boundary conditions
\begin{equation}
\theta^{\prime}(0) = \theta^{\prime}(\pm L/2)=0  \label{bc}
\end{equation}

The Lagrange multiplier $\omega$ ensures the condition of a given
normalized magnetization
\begin{equation}
n_z = \int\limits_{-L/2}^{L/2}\cos\theta\frac{\mathrm{d}x}{L}
\label{MGP}
\end{equation}

Simple analysis shows that for any given values of $n_z$ and $L$
there is a critical value of parameter $\alpha=\alpha_0(n_z,L)$ so
that at $\alpha<\alpha_0$ Eq.(\ref{eqXXZ}) has a single and uniform
solution $\theta(x)=\theta_0=\arccos(n_z)$. The energy of this
solution is
\begin{equation}
E = Ls^2\alpha (1-n_z^2)  \label{Elin}
\end{equation}

At $\alpha>\alpha_0$ another, non-trivial solution of
Eq.(\ref{eqXXZ}) appears, which is a precursor of localized
soliton solution. In particular, for the case $n_z=0$ the critical
value is $\alpha_0=2\pi^2/L^2$ and the uniform solution at
$\alpha<\alpha_0$ is $\theta_0=\pi/2$. The non-trivial solution at
$\alpha\gtrsim \alpha_0$ can be expanded in small parameter
$\vartheta=\sqrt{(\alpha-\alpha_0)/2\alpha_0}\ll 1$:
\begin{equation}
\theta (\xi )=\pi /2+2\vartheta \cos \xi +2\vartheta ^{3}\xi \sin
\xi -\frac{\vartheta ^{3}}{6}\cos (3\xi )
\end{equation}
where $\xi=(2\alpha)^{1/2}x$.

The energy for this case can be written as
\begin{eqnarray}
\frac{E}{s^{2}\sqrt{\alpha }} &=&l,\qquad l<l_{0} \\
\frac{E}{s^{2}\sqrt{\alpha }} &=&l-\frac{\sqrt{2}(l-l_0)^2}{\pi
},\qquad l\gtrsim l_0
\end{eqnarray}
where $l=L\sqrt{\alpha}$ and $l_0=\pi\sqrt{2}$. Therefore, the
second derivative of $E$ with respect to $l$ is discontinuous at
$l=l_0$ ($\alpha=\alpha_0$).

At $l\to\infty $ the solution for the case $n_z=0$ becomes
\begin{equation}
\tan \left( \theta /2\right) =\exp \left[ \sqrt{2\alpha }(x+L/4)\right]
\end{equation}
and $E$ tends to the known results $4s^2\sqrt{2\alpha}$.

A similar behavior of the soliton solution takes place for other
values of $\lambda$. It is straightforward to obtain the value
$\alpha_0$ and the soliton solution near $\alpha_0$ for the
general case in the same manner as for the case $\lambda=0$. For
example, for $\lambda=1/4$ and $n_z=0$ the critical value is
$\alpha_0=2\pi^4/L^4$ and the solution at $\alpha\gtrsim\alpha_0$
is
\begin{equation}
\theta(\xi)= \frac{\pi}{2} +\vartheta \cos\xi
+\frac{\vartheta^3}{2}\xi \sin\xi
+\frac{\vartheta^3}{60}\cos(3\xi)
\end{equation}
where $\xi=(8\alpha)^{1/4}x$.

The energy in this case is
\begin{eqnarray}
\frac{E}{s^2\alpha^{3/4}} &=&l,\qquad l<l_0  \nonumber \\
\frac{E}{s^2\alpha^{3/4}} &=&l-\frac{3(l-l_0)^2}{2^{1/4}\pi},\qquad
l\gtrsim l_0
\end{eqnarray}
where $l=\alpha^{1/4}L$ and $l_0=2^{1/4}\pi$.

\begin{figure}[tbp]
\includegraphics[width=3in]{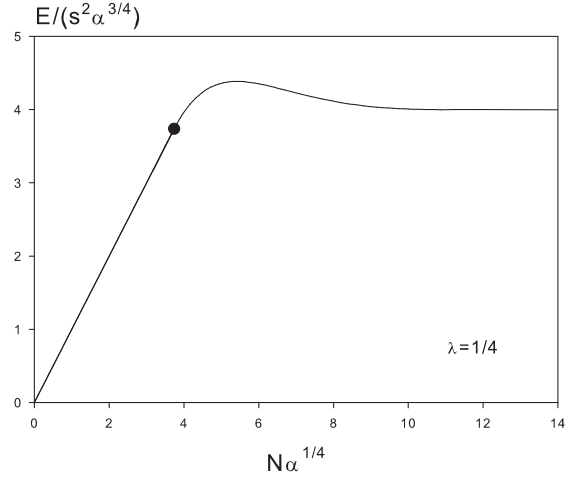}
\caption{Scaled soliton energy versus finite-size parameter
$N\alpha^{1/4}$ for $\lambda=1/4$ and $k=0$. Circle is a critical
point of soliton creation.} \label{GP}
\end{figure}

The calculated dependence $E(l)$ for $\lambda=1/4$ and $n_z=0$ is
shown in Fig.\ref{GP}. As one can see the energy rapidly saturates
to its asymptotic value at $l>l_0$. Therefore, in order to reduce
finite-size effects in numerical calculations near the transition
point $\lambda=1/4$ one should choose the system size $N$ so that
the parameter $\alpha^{1/4}N>10$.

The critical $\alpha_0$ for a given magnetization $n_z$ can be
restored from the case $n_z=0$ by the relation
\begin{equation}
\alpha_0(n_z,L)=\frac{\alpha_0(0,L)}{1-n_z^2}
\end{equation}

The process of the formation of the kink (domain-wall) occurs in a
somewhat different way. The kink is realized in open chain with the
boundary conditions $\theta(-L/2)=0$, $\theta(L/2)=\pi$. In this
case the lowest state at small $\alpha$ is the spiral one
\begin{equation}
\theta(x)= \frac{\pi}{2} + \frac{\pi x}{L}
\end{equation}
and its energy remains finite at $\alpha=0$ in contrast with the
solitons. For example, for $\lambda=1/4$ it is
\begin{equation}
E = s^2\frac{2\pi^4}{L^3} + s^2\frac{\alpha L}{2}
\end{equation}

When $\alpha$ increases the spiral configuration continuously
transforms to the kink. In this respect the situation is different
from the definite and sharp transition from the uniform to
soliton-like state studied before. The crossover between the
spiral and the kink states takes place at $l\sim 1$. At
$l\to\infty$ the energy of the kink is a half energy of the
soliton.

It is worth noting that the transition of such kind coming from the
finite-size effects occurs in the quantum spin model (\ref{HH}) as
well. For very small $\alpha$, when the corresponding parameter
$l\ll 1$, the system is in a uniform state, which is the state with
a given total projection $S^z$ and the maximal total spin
$S=S_{\max}=sN$. The energy of this state is given by
Eq.(\ref{Elin}) with $n_z=S^z/sN$ as in the classical model. This
uniform state transforms continuously to the localized state of
bound magnon complex when $\alpha$ increases. The crossover between
the uniform and the localized states takes place at
$\alpha^{1/2}N\sim 1$ for $\lambda=0$ and at $\alpha^{1/4}N\sim 1$
for $\lambda=1/4$.

\section{Quantum model. Mapping to $\delta$-attractive bose model.}

It is known that there is a resemblance between the classical
solitons and the quantum magnon complexes at $\lambda=0$
\cite{Schneider}. In the preceding sections we observed such a
resemblance for the classical F-AF model and its quantum
counterpart with $s=1/2$ as well. Here we consider the excitations
in the quantum model with general $s$ and compare them with the
classical solitons. A standard method to treat the quantum spin
models is a mapping of the spin-Hamiltonian to the Bose-one.
Though this method is approximate it gives important indications
about the behavior of the spin systems. Using the Dyson-Maleev
transformation
\begin{eqnarray}
S_i^z &=&a_i^+a_i-s  \nonumber \\
S_i^+ &=&\sqrt{2s}a_i^+\left( 1-\frac{a_i^+a_i}{2s}\right)
\nonumber \\
S_i^- &=&\sqrt{2s}a_i
\end{eqnarray}
we represent the Hamiltonian (\ref{HH}) in the form
\begin{equation}
H = \sum_k \varepsilon_k a_k^+ a_k +\frac{1}{N} \sum_{k_1,k_2,q}
V(k_1,k_2,q) a_{q+k_1}^+ a_{q-k_1}^+ a_{q+k_2} a_{q-k_2}
\label{Hbose}
\end{equation}
where $a_k$ and $a_k^+$ are conventional bose-operators.

In the long-wavelength and weakly-coupling limit
$V(k_1,k_2,q)=-\alpha$ and the one-magnon spectrum $\varepsilon_k$
in the vicinity of the IT point is
\begin{equation}
\varepsilon_k = 2s\alpha -s\gamma k^2 +\frac{sk^4}{4}
\end{equation}

In this limit Hamiltonian (\ref{Hbose}) is equivalent to the
Bose-model with the attractive $\delta$ -function interaction
\begin{equation}
H=2sm\alpha +s\sum_{i=1}^{m}\left( \gamma \frac{\partial ^{2}}{\partial
x_{i}^{2}}+\frac{1}{4}\frac{\partial ^{4}}{\partial x_{i}^{4}}\right)
-2\alpha \sum_{j>i=1}^{m}\delta (x_{i}-x_{j})  \label{Hbose2}
\end{equation}
where $m$ denotes number of the Bose particles. For the case
$\lambda=0$ one can neglect the fourth-order derivative term and
such model becomes exactly solvable one \cite{Guire}. However, in
the vicinity of the IT point the fourth-order derivative term in
Eq.(\ref{Hbose2}) is important and it destroys the exact solution
found in Ref.\cite{Guire}. Therefore, we have to use approximations.

First, we consider the two-boson bound state near the IT point and
compare it with the two-magnon bound state found in Ref.\cite{DK09}.
The wave function of two bosons with the total momentum $2k$ and the
energy $E_2$ is $\Psi(x_1,x_2)=e^{ik(x_2+x_1)}\psi(x_2-x_1)$. The
function $\psi$ is determined from the Schr\"{o}dinger equation
\begin{equation}
(2s\gamma -3sk^{2})\frac{\partial ^{2}\psi (x)}{\partial
x^{2}}+\frac{s}{2} \frac{\partial ^{4}\psi (x)}{\partial
x^{4}}-2\alpha \delta (x)\psi (x)=\varepsilon _{b}\psi (x)
\end{equation}
where $x=x_2-x_1$ and $\varepsilon_b = 2\varepsilon_k-E_2$
($\varepsilon_b>0$) is the binding energy. Here we see that the
fourth-order derivative term introduces the dependence of the
function $\psi(x)$ and the binding energy $\varepsilon_b$ on the
total momentum.

The function $\psi(x)$ as well as its first- and second-order
derivatives are continuous at $x=0$, while the third-order
derivative satisfies the condition
\begin{equation}
\frac{s}{4}\left. \frac{\partial^3\psi}{\partial x^3}\right\vert
_{-0}^{+0} = \alpha\psi(0)
\end{equation}

The wave function of the bound state of two particles with $\delta$
-attraction has a form:
\begin{equation}
\psi (x)= e^{-\kappa |x|}\left[ \eta \cos (\eta x)+\kappa \sin (\eta
|x|) \right]  \label{psi2boson}
\end{equation}
where the parameters $\kappa$ and $\eta$ are determined by the above
stated conditions for the function $\psi(x)$ and its derivatives at
$x=0$. The solution has a form
\begin{eqnarray}
\kappa &=&\left( \frac{\alpha }{4s}\right) ^{1/3}g(\beta )
\nonumber \\
\eta &=&\left(\frac{\alpha}{4s}\right)^{1/3}\sqrt{8\beta
+g^2(\beta)}
\end{eqnarray}
where
\begin{eqnarray}
\beta &=&\frac{\gamma -3k^{2}/2}{\left( 2\alpha /s\right) ^{2/3}} \\
g(\beta ) &=&\left( 1+\sqrt{1+(4\beta /3)^{3}}\right)
^{1/3}-\frac{4\beta /3}{\left( 1+\sqrt{1+(4\beta /3)^{3}}\right)
^{1/3}}
\end{eqnarray}

The binding energy is
\begin{equation}
\varepsilon_b=\frac{2^{1/3}\alpha ^{4/3}}{s^{1/3}g^{2}(\beta )}
\end{equation}

In particular, for $\gamma=0$ ($\lambda=1/4$) and $k=0$ it is
\begin{equation}
\varepsilon_b = \frac{\alpha ^{4/3}}{(2s)^{1/3}} \label{E2boson}
\end{equation}

The wave function (\ref{psi2boson}) has one interesting feature.
It oscillates when $\eta$ is real and has a form of superposition
of two decaying exponents otherwise:
$\psi(x)=e^{-\kappa_1|x|}+ce^{-\kappa_2|x|}$. The boundary between
these regions is defined by equation $\eta=0 $, which has a
solution $\beta=-1/2^{5/3}$. Thus, the wave function of two
coupled bosons oscillates for $2\gamma>3k^2-(\alpha/s)^{2/3}$ and
does not oscillate for $2\gamma\leq 3k^2-(\alpha/s)^{2/3}$. The
oscillation region diminishes with the increase of the total
momentum. Certainly, the oscillatory behavior of the wave function
is the effect of the frustration similarly to that observed for
the solitons in the spin model.

We can compare Eq.(\ref{E2boson}) with the binding energy of two
magnons for $s=1/2$ found by us in \cite{DK09}
\begin{equation}
E_{b}=\alpha ^{4/3}-\frac{2}{3}\alpha ^{5/3}+O(\alpha ^{2})  \label{E2magnon}
\end{equation}

It follows from Eqs.(\ref{E2boson}) and (\ref{E2magnon}) that the
leading term in $E_b$ at $\alpha\to 0$ coincides up to numerical
factor with the two-boson binding energy, but subsequent terms in
the expansion of $E_b$ are absent in (\ref{E2boson}). Similar
difference between $E_b$ and $\varepsilon$ takes place for any
values of $\lambda$. For example, for $\lambda=0$ it is
\begin{eqnarray}
\varepsilon_b &=&\frac{\alpha ^{2}}{2s}  \nonumber \\
E_b &=&\frac{\alpha ^{2}}{2s}-\frac{4s-1}{8s^{3}}\alpha
^{3}+O(\alpha ^{4})
\end{eqnarray}

As will be shown below, this difference plays a key role at
comparison of the multimagnon and multiboson bound states.

The exact wave function of the $m$-boson state at $\lambda=0$ is
known for any $m$ \cite{Guire}. As was shown in
Ref.\cite{Calogero} at $\lambda=0$ the Hartree approximation
correctly reproduces the exact energy of the multiboson states
($m\gg 1$) with zero total momentum. Therefore, we expect that
this approach gives reliable results for $m\gg 1$ when
$\lambda\neq 0$. Here we consider Bose model (\ref{Hbose2}) in a
parametric regime corresponding to the vicinity of the IT point of
model (\ref{HH}).

The energy functional for zero total momentum in the Hartree
approximation is
\begin{equation}
E_m = 2sm\alpha +m\int_{-\infty}^{\infty}\mathrm{d}x\left[
-s\gamma\left(\frac{\partial\phi}{\partial x}\right)^2
+\frac{s}{4}\left(\frac{\partial^2\phi}{\partial x^2}\right)^2
-m\alpha\phi^4\right] \label{Ehartree}
\end{equation}

Rescaling $\xi=(m\alpha/s)^{1/3}x$ and
$\phi(\xi)=(m\alpha/s)^{1/6}\chi(\xi)$ transforms the energy
functional (\ref{Ehartree}) to
\begin{equation}
E_m = 2sm\alpha +sm(\frac{m\alpha}{s})^{4/3}
\int_{-\infty}^{\infty} \mathrm{d}\xi
\left[-\beta\left(\frac{\partial\chi}{\partial\xi}\right)^2
+\frac{1}{4}\left(\frac{\partial^2\chi}{\partial\xi^2}\right)^2
-\chi^4(\xi)\right] \label{Eharnor}
\end{equation}
where $\beta=\gamma/(m\alpha/s)^{2/3}$. The function $\phi(x)$
satisfies the conventional normalization condition, which after
rescaling gives the same normalization condition for $\chi(\xi)$
\begin{equation}
\int_{-\infty}^{\infty}\phi^2(x)\mathrm{d}x =
\int_{-\infty}^{\infty}\chi^2(\xi)\mathrm{d}\xi =1 \label{norma}
\end{equation}

The Hartree equation comes from the minimization of the energy
over $\chi(\xi)$. For $m\gg 1$ it has a form:
\begin{equation}
\beta\frac{\partial^2\chi}{\partial\xi^2}
+\frac{1}{4}\frac{\partial^4\chi}{\partial\xi^4} -2\chi
^{3}=\epsilon \chi \label{eqhartree}
\end{equation}
where $\epsilon$ is the Lagrange multiplier secured the norma
condition (\ref{norma}).

Unfortunately, the solution of this equation at present is unknown.
However, if we assume the existence of a localized solution of
Eq.(\ref{eqhartree}), then the integral in Eq.(\ref{Eharnor})
converges yielding some function of parameter $\beta$. So, the
energy of $m$-boson bound complex takes the form
\begin{equation}
E_m = 2sm\alpha -sm(\frac{m\alpha}{s})^{4/3}G_b(\beta)
\label{Eboson}
\end{equation}

The behavior of the function $G_b(\beta)$ obtained by the
numerical minimization of the functional (\ref{Eharnor}) is
demonstrated in Fig.\ref{Gbeta}. Numerical calculations also
showed that there is a critical value of the parameter
$\beta\approx -0.57$ separating the regions with ($\beta>-0.57$)
and without ($\beta<-0.57$) oscillations of the function
$\phi(x)$. The value $\beta=-0.57$ is the same as was found for
the classical spin model, which certificates the distinct
correspondence between the boson and the classical spin models for
small solitons. However, the bound complex of bosons exists in the
whole region of parameter $\beta$, which means that the
instability of solitons studied in Sec.III comes from the effects
neglecting in the boson approach.

It is interesting to compare Eq.(\ref{Eboson}) with the energy of
the $m$-magnon bound state of the quantum model (\ref{HH}) with
$s=1/2$ at $\lambda=1/4$. The latter has been obtained in
Ref.\cite{DK09} and has a form
\begin{equation}
E_{m} = m\alpha -C_1m^{7/3}\alpha ^{4/3}+C_2m^{11/3}\alpha^{5/3}
+\ldots
\end{equation}
where $C_1$ and $C_2$ are numerical coefficients.

Similarly to the case $m=2$ the leading terms of the expansions of
these two energies in small $\alpha$ coincide with
Eq.(\ref{Eboson}) up to numerical factors. However, energy
(\ref{Eboson}) tends to $-\infty$ at $m\to\infty$ (collapse
phenomenon), while the energy of $m$-magnon bound complex in the
spin model are finite at $m\to\infty$ and behaves as
$E_m\sim\alpha^{3/4}$ \cite{DK09}. Therefore, subsequent terms in
the energy expansion in $\alpha$ of the spin model are responsible
for a short distance repulsion of magnons preventing the collapse.

A comparison of the expression for the energy of the classical
soliton of size $M$ (\ref{Escal}) with a formula for the energy of
$m$-boson complex (\ref{Eboson}) indicates that the obtained
expression for the energy of $m$-boson complex (\ref{Eboson}) is a
particular case of a more general scaling relation (\ref{Escal})
for $k=0$ and
\begin{equation}
f_b(\mu,\nu,0) = 2\nu -\nu^{7/3}G_b\left(\frac{\mu}{\nu^{2/3}}
\right) \label{fboson}
\end{equation}

As we discussed in Sec.IV the scaling function of the boson model
$f_b(\mu,\nu,0)$ at $\nu\ll 1$ coincides with the scaling function
of the classical spin model. The comparison of this equation with
Eq.(\ref{Escal}) allows us to assume that Eq.(\ref{fboson})
represents two first terms of the scaling function of the quantum
spin-$s$ model (including the classical limit) in small parameter
$\nu$
\begin{equation}
f_s(\mu,\nu,0) =\sum_{n=0}^{\infty}
\nu^{4n/3+1}g_{n,s}\left(\frac{\mu }{\nu ^{2/3}}\right)
\label{fseries}
\end{equation}
where functions $g_{n,s}$ depend on $s$.

Eq.(\ref{fboson}) reproduces (probably exactly for any $s$) two
first terms in this expansion. But the expansion (\ref{fseries})
contains infinite number of terms in contrast with
Eq.(\ref{fboson}). It leads to the finite energy of both the
classic solitons and the quantum bound magnon complexes while the
energy of $m$-boson complex diverges at $m\to\infty$.

Summarizing all the above facts, we believe that the bound energy
of $m$ -magnon complex for a quantum spin-$s$ model (\ref{HH})
near the IT point is correctly described by the classical scaling
formula (\ref{Escal})
\begin{equation}
E = s^2\alpha^{3/4}f_s(\mu,\nu,k)  \label{Es}
\end{equation}

The scaling functions $f_s$ for quantum spin-$s$ case does not
coincide with the function $f$ obtained in the classical continuum
approach, though all of them have very similar behavior and
$\lim\limits_{s\to\infty}f_s=f$.

\section{Conclusion}

We studied the soliton excitations in the classical F-AF model with
the easy-axis anisotropy. The F-AF model has two parameters: the
frustration parameter $\lambda=|J_2|/J_1$ and the anisotropy
$\alpha$. We found that in a weakly anisotropic limit ($\alpha\ll
1$) the behavior of the soliton solutions for small frustration
parameter $\lambda<1/4$ is qualitatively similar to that for the
exactly solvable easy-axis XXZ chain ($\lambda=0$ case). However,
the situation drastically changes near the IT point ($\alpha=0$,
$\lambda=1/4$), where the transition from the ferromagnetic to the
spiral ground state takes place in the isotropic case. In the
vicinity of the IT point the corresponding energy functional in the
continuum approximation qualitatively changes and does not admit the
exact solution. The analysis of the derived energy functional
allowed us to estimate the behavior of the transition line between
the ferromagnetic to the spiral ground state in ($\alpha,\lambda$)
plane near the IT point.

We mainly interested in the behavior of the solitons in the vicinity
of the IT point. We showed that these localized states are separated
from the ferromagnetic state by a finite gap. The dependence of the
soliton energy (the gap) on model parameters near the IT point was
established on a base of the scaling arguments. As a result we found
that the soliton energy is proportional to $\sim\alpha^{3/4}$ and is
expressed by the scaling function (\ref{Escal}) depending on three
scaling parameters: $\mu=(\lambda-1/4)/\sqrt{\alpha}$, the scaled
soliton size $\nu$ and the momentum $k$.

The analysis of the asymptotic solutions of the corresponding
equation of motion provided us with the necessary conditions of the
soliton stability. It was shown that solitons of all sizes and any
momentum exist when $\mu\leq 0$, while in the region $\mu>0$ there
are definite restrictions on the soliton size $\nu$ and the momentum
$k$. The distribution of the allowable values of the soliton
parameters in ($\nu,k$) plane for $\mu>0$ has very complicated form,
which can be determined numerically only. For example, the static
solitons ($k=0$) only of the middle size $3.5<\nu<8$ are stable at
$\mu=1.5$. Nevertheless, some facts about the soliton existence
region was ascertained analytically and then confirmed by numerical
calculations. In particular, small static ($k=0$) solitons are
unstable for $\mu>0$, while small solitons with non-zero momentum
exist for any $\mu$. Large solitons ($\nu\gg 1$) exist for
$\mu<\sqrt{2}$ and, therefore, they survive in the part of the
spiral phase $1\lesssim\mu<\sqrt{2}$. Though the soliton excitations
in the spiral phase lie in the high-energy part of the spectrum,
they can play an essential role in the magnetization processes.

In the isotropic limit ($\mu\to\infty$) only small solitons with
non-zero momentum survive and the maximal allowable size of these
solitons increases when the frustration parameter tends to the
critical value $\lambda\to 1/4$. Such a dependence of the soliton
size on the frustration parameter is qualitatively similar to the
observation that the size of the multimagnon bound complexes with
$k=\pi$ in the quantum F-AF model grows when $\lambda\to 1/4$.

We found that the frustration reveals itself in the oscillating
shape of solitons. The amplitude of the oscillations grows at
approaching to the spiral phase and further inside of it. Generally,
at some value of $\mu$ the solution starts to oscillate without the
decay at infinity, i.e., soliton-like solution with given values of
$\nu$ and $k$ disappears.

We studied the finite-size effects on the soliton solutions. It was
shown that the soliton solution on finite ring originates in the
uniform (non-localized) state. The transition from the uniform to
the soliton-like state occurs at the critical value of the
anisotropy $\alpha_0(L)\sim L^{-4}$,  below which the uniform
solution is realized. This finite-size effect is similar to the
well-known Gross-Pitaevskii transition in the Bose-systems.

In order to establish a connection between the properties of the
solitons and the multi-magnon complexes of the quantum counterpart
of this model we used the Dyson-Maleev mapping of the spin model to
the Bose one. It turned out that the dependence of the energy of
boson bound complexes on model parameters represents a particular
case of the found scaling expression for the classical soliton
energy. Moreover, the energy of the bound magnon complexes for
quantum spin-$1/2$ model found by us before perfectly coincides with
the scaling equation for soliton energy. Therefore, we believe that
the bound energy of multi-magnon complex for a quantum spin-$s$
model near the IT point is characterized by the same critical
exponents and by the identical scaling parameters as the soliton
energy, though the corresponding scaling functions are different for
different $s$. It is known that such a resemblance takes place for
the Heisenberg ferromagnetic chain with an easy-axis anisotropy. Our
study shows that the multi-magnon complexes behave substantially as
the classical objects for the frustrated model as well.

For future it would be interesting to study the behavior of the
classical spin model deeply in the spiral phase. We believe that it
can help in understanding of unusual magnetization processes and
shed light on the behavior of the multi-magnon complexes at high
magnetic fields.


\begin{thebibliography}{99}

\bibitem{review} H.-J.~Mikeska and A. K.~Kolezhuk, in \textit{Quantum
Magnetism}, Lecture Notes in Physics Vol. \textbf{645}, edited by
U.~Schollw\"{o}ck, J.~Richter, D. J. J.~Farnell, and R. F.~Bishop,
Eds. (Springer-Verlag, Berlin, 2004), p. 1.

\bibitem{Mizuno} Y.~Mizuno, T.~Tohyama, S.~Maekawa, T.~Osafune, N.~Motoyama,
H.~Eisaki, and S.~Uchida, Phys. Rev. B \textbf{57}, 5326 (1998).

\bibitem{Masuda} T.~Masuda, A.~Zheludev, A.~Bush, M.~Markika, and
A.~Vasiliev, Phys. Rev. Lett. \textbf{92}, 177201 (2004).

\bibitem{Hase} M.~Hase, H.~Kuroe, K.~Ozawa, O.~Suzuki, H.~Kitazawa, G.~Kido
and T.~Sekine, Phys. Rev. B \textbf{70}, 104426 (2004).

\bibitem{stefan05} S.-L.~Drechsler, J.~Malek, J.~Richter, A.~S.~Moskvin,
A.~A.~Gippius, and H.~Rosner, Phys. Rev. Lett. \textbf{94}, 039705
(2005).

\bibitem{Capogna} L.~Capogna, M.~Mayr, P.~Horsch, M.~Raichle, R.~K.~Kremer, M.~Sofin,
A.~Maljuk, M.~Jansen, and B.~Keimer, Phys. Rev. B \textbf{71},
140402 (2005).

\bibitem{stefan08} J.~Malek, S.-L.~Drechsler, U.~Nitzsche, H.~Rosner, and
H.~Eschrig, Phys. Rev. B \textbf{78}, 060508(R) (2008).

\bibitem{Chubukov} A. V.~Chubukov, Phys. Rev. B \textbf{44}, 4693 (1991).

\bibitem{Itoi} C.~Itoi and S.~Qin, Phys. Rev. B \textbf{63}, 224423 (2001).

\bibitem{KO} V.~Ya.~Krivnov and A.~A.~Ovchinnikov, Phys. Rev. B \textbf{53},
6435 (1996).

\bibitem{DK06} D.~V.~Dmitriev and V.~Ya.~Krivnov, Phys. Rev. B \textbf{73},
024402 (2006).

\bibitem{Vekua} F.~Heidrich-Meisner, A.~Honecker, and T.~Vekua, Phys. Rev. B
\textbf{74}, 020403(R) (2006).

\bibitem{Lu} H.~T.~Lu, Y. J.~Wang, S.~Qin, and T.~Xiang, Phys. Rev. B
\textbf{74}, 134425 (2006).

\bibitem{DKR} D.~V.~Dmitriev, V.~Ya.~Krivnov, and J.~Richter, Phys. Rev. B
\textbf{75}, 014424 (2007).

\bibitem{Kecke} L.~Kecke, T.~Momoi, and A.~Furusaki, Phys. Rev. B
\textbf{76}, 060407(R) (2007); T.~Hikihara, L.~Kecke, T.~Momoi, and
A.~Furusaki, Phys. Rev. B \textbf{78}, 144404 (2008).

\bibitem{Laeuchli} A.~M.~Laeuchli, J.~Sudan, and A.~Luscher, J.
Phys. Conf. Series \textbf{145}, 012057 (2009).

\bibitem{Somma} R.~D.~Somma and A.~A.~Aligia, Phys. Rev. B \textbf{64},
024410 (2001); R.~Jafari and A.~Langari, Phys. Rev. B \textbf{76},
014412 (2007); A.~Avella, F.~Mancini, and E.~Plekhanov, Eur. Phys.
J.: Cond. Matter, \textbf{66}, 295 (2008).

\bibitem{DK08} D.~V.~Dmitriev and V.~Ya.~Krivnov, Phys. Rev. B \textbf{77},
024401 (2008).

\bibitem{DK09} D.~V.~Dmitriev and V.~Ya.~Krivnov, Phys. Rev. B \textbf{79},
054421 (2009).

\bibitem{Johnson} J.~D.~Johnson and J.~C.~Bonner, Phys. Rev. B \textbf{22},
251 (1980).

\bibitem{Schneider} T.~Schneider, Phys. Rev. B \textbf{24}, 5327 (1981).

\bibitem{Steiner} H.-J.~Mikeska and M.~Steiner, Adv. Phys. \textbf{40},
191 (1991).

\bibitem{KIK} A.~M.~Kosevich, B.~A.~Ivanov and A.~S.~Kovalev, Phys. Rep.
\textbf{194}, 117 (1990).

\bibitem{Laksh} M.~Lakshmanan, Phys. Lett. A \textbf{61}, 53 (1977); L. A.
Takhtajan, Phys. Lett. A \textbf{64}, 235 (1977).

\bibitem{Ovchinnikov} A.~A.~Ovchinnikov, JETP Lett. \textbf{5}, 38
(1967); I.~G.~Gochev, JETP \textbf{34}, 892 (1972).

\bibitem{gochev} I.~G.~Gochev, JETP \textbf{58}, 115 (1983).

\bibitem{ivanov} B.~A.~Ivanov, A.~Yu.~Merkulov, V.~A.~Stephanovich, and
C.~E.~Zaspel, Phys. Rev. B \textbf{74}, 224422 (2006).

\bibitem{Kosevich} B.~A.~Ivanov and A.~M.~Kosevich, Sov. J. Low Temp. Phys.
\textbf{9}, 439 (1983).

\bibitem{Mikeska} C.~Etrich, H.-J.~Mikeska, E.~Magyari, H.~Thomas, and
R.~Weber, Z. Phys. B \textbf{62}, 97 (1985).

\bibitem{IM} B.~A.~Ivanov and H.-J.~Mikeska, Phys. Rev. B \textbf{70},
174409 (2004).

\bibitem{KosJETP} A.~M.~Kosevich, B.~A.~Ivanov, and A.~S.~Kovalev, JETP Lett.
\textbf{25}, 486 (1977).

\bibitem{BEGP} R.~Kanamoto, H.~Saito, and M.~Ueda, Phys. Rev. A \textbf{67},
013608 (2003); G.~M.~Kavoulakis, Phys. Rev. A \textbf{67}, 011601(R)
(2003); K.~Sakmann, A.~I.~Streltsov, O.~E.~Alon, and
L.~S.~Cederbaum, Phys. Rev. A \textbf{72}, 033613 (2003).

\bibitem{Guire} J.~B.~McGuire, J. Math. Phys. \textbf{5}, 622 (1964).

\bibitem{Calogero} F.~Calogero and A.~Degasperis, Phys. Rev. A \textbf{11},
265 (1975).

\end{thebibliography}
\end{document}